\documentclass[aps,pre,superscriptaddress,amsmath,amssymb,twocolumn]{revtex4}
\usepackage{graphicx}
\usepackage{amsmath}
\usepackage{hyperref}
\usepackage{color}
\definecolor{darkblue}{RGB}{8,81,156}
\definecolor{dark-red}{RGB}{178,24,43}
\hypersetup{colorlinks,breaklinks,
            linkcolor=darkblue,urlcolor=darkblue,
           anchorcolor=darkblue,citecolor=darkblue}

\date{\today}

\allowdisplaybreaks

\begin{document}

\title{Perspective: Surface Freezing in Water: A Nexus of Experiments and Simulations}

\author{Amir Haji-Akbari}
\email{amir.hajiakbaribalou@yale.edu}
\affiliation{Department of Chemical and Environmental Engineering, Yale University, New Haven, CT  06520, \textcolor{black}{USA}}

\author{Pablo G. Debenedetti}
\email{pdebene@exchange.princeton.edu}
\affiliation{Department of Chemical and Biological Engineering, Princeton University, Princeton, NJ 08540, \textcolor{black}{USA}}

\date{\today}

\begin{abstract}
Surface freezing is a phenomenon in which crystallization is enhanced at a vapor-liquid interface. In some systems, such as  $n$-alkanes, this enhancement is dramatic, and results in the formation of a crystalline layer at the free interface even at temperatures slightly above the equilibrium bulk freezing temperature. There are, however, systems in which the enhancement is purely kinetic, and only involves faster nucleation  at or near the interface. The first, thermodynamic, type of surface freezing is easier to confirm in experiments, requiring only the verification of the existence of crystalline order at the interface. The second, kinetic, type of surface freezing is far more difficult to prove experimentally. One material that is suspected of undergoing the second type of surface freezing is liquid water. Despite strong indications that the freezing of liquid water is kinetically enhanced at vapor-liquid interfaces, the findings are far from conclusive, and the topic remains controversial. In this perspective, we present a simple thermodynamic framework to understand conceptually and distinguish these two types of surface freezing. We then briefly survey fifteen years of experimental and computational work aimed at elucidating the surface freezing conundrum in water. 
\end{abstract}

\maketitle

\section{Introduction\label{section:intro}}

Confinement can alter the way in which materials behave, from their optoelectronic~\cite{MakPRL2010, SugawaACSNano2015}, mechanical~\cite{GreerActaMaterialia2005, ChengNanoLett2014} and transport~\cite{BergmanNature2000, ZhangPRL2006, ZhuEdigerPRL2011, ShiJCP2011, EdigerMacromolecules2014, HajiAkbariJCP2014, HajiAkbariJCP2015, ZhangPNAS2017} properties, to their phase diagrams~\cite{McKennaJCP1990, CiccoPRL1998, AlbaSimionescoJPhysCondMat2006, JineshPRL2008, ZhangPNAS2017} and phase transition kinetics~\cite{ZhangPRL2006, JineshPRL2008, FalahatiCurrBiol2016, AltabetPNAS2017, FalahatiPNAS2017}. Appreciable changes in physical properties, however, require confinement in geometries characterized by very small length scales, which, in the present context, means that the characteristic length scale of the confining geometry, $l_c$, is comparable to that of  intermolecular interactions, $l_i$. One major exception, however, is the kinetics of first-order phase transitions, which can be dramatically affected even when $l_c$ is orders of magnitude larger than $l_i$. This is because of the activated nature of first-order phase transitions, which typically occur through a nucleation and growth mechanism. In the nucleation-limited regime, even a modest decrease in the height of the nucleation barrier due to the presence of an interface can result in dramatically higher rates, which will in turn affect the macroscopic behavior of the system at length scales much larger than $l_i$.

Among first-order transitions, disorder-order transitions, such as crystallization, are particularly sensitive to the presence of interfaces, because homogeneous nucleation of phases with long-range spatial order usually requires overcoming very large nucleation barriers, and only becomes likely at temperatures considerably lower than $T_m^{\text{bulk}}$, the equilibrium bulk melting temperature. As a result, in most materials, freezing occurs heterogeneously, in the presence of an exogenous solid interface. Free (i.e.,~vapor-liquid) interfaces can also facilitate freezing, and such a possibility can be very important in a variety of scientifically and technologically important phenomena, such as ice formation in the atmosphere, or crystallization of metallic glasses. In some materials, surface freezing can be very dramatic, and can occur even at temperatures above $T_{m}^{\text{bulk}}$.  For instance, a frozen monolayer of the $R_{\text{II}}$ rotator phase emerges at free interfaces of $n$-alkanes at temperatures as large as 3~K higher than $T_{m}^{\text{bulk}}$~\cite{WuScience1993, WuPRL1995, SeflerChemPhysLett1995,  PfohlChemPhysLett1996,  OckoPRE1997}. Such frozen monolayers have also been observed in other chain molecules, including dry and hydrated linear alcohols~\cite{SeflerChemPhysLett1995, GangPhysRevE1998}.  Frozen monolayers have also been observed in some alloys, such as Au-Si~\cite{ShpyrkoScience2006, ShpyrkoPhysRevB2007}, Ga-Bi~\cite{TurchaninPCCP2002} and Ga-Pd~\cite{TurchaninPCCP2002, FreylandJPhysCondMat2002}. A related phenomenon, known as surface ordering, occurs in materials that form liquid crystals, where, at temperatures above $T_{o}^{\text{bulk}}$, the bulk isotropic-to-liquid crystalline transition temperature, a liquid crystalline layer forms at the vapor-liquid interface~\cite{SwansonPRL1989}.

When surface freezing occurs  at temperatures above $T_m^{\text{bulk}}$, it can  be easily observed and characterized in experiments since this involves simply detecting crystalline order at the surface of a liquid slightly above $T_m^{\text{bulk}}$. Experimental accounts of such thermodynamic manifestations of surface freezing are therefore rarely controversial. There is, however, a more subtle type of surface freezing that can only occur at temperatures below $T_m^{\text{bulk}}$, and that involves an enhancement of nucleation kinetics at a vapor-liquid interface. It is far more difficult to probe this second type of surface freezing experimentally, and its accounts in the literature are often controversial. Unambiguous detection and quantification issues notwithstanding, surface-facilitated nucleation can dramatically impact the kinetics of crystallization in confined systems, such as  droplets and thin films. Therefore, determining whether crystallization of a particular material is enhanced near free interfaces is crucial in understanding and engineering the behavior of systems that include confined states of such a material.

In this perspective, our main focus will be on water, which is suspected of undergoing this second, kinetic, type of surface freezing~\cite{TabazadehPNAS2002}. An important example of surface freezing in water involves ice formation in atmospheric processes, as the amount of ice in a cloud affects its light-absorbing properties and its propensity to produce rain and snow~\cite{DeMottPNAS2010}. Whether ice nucleation is enhanced or suppressed at a vapor-liquid interface is consequential in determining the behavior of atmospheric clouds, which are comprised of polydisperse water microdroplets. The large surface area of such droplets makes the spatiotemporal distribution of freezing events in a cloud highly sensitive to free surface-enhanced nucleation. Despite some strong supporting evidence, the question of whether vapor-liquid interfaces accelerate ice nucleation in water is  controversial, and is considered one of the ten most important open questions regarding ice and snow~\cite{RauschNature2013}. This perspective is aimed at addressing this question and is organized as follows. In Section~\ref{section:thermo}, we discuss the thermodynamics of surface freezing.  Section~\ref{section:experimental} provides a brief overview of the different experimental approaches used to detect surface freezing. In Section~\ref{section:water}, we present a summary of fifteen years of experimental and computational work devoted to resolving the surface freezing conundrum in water. And Section~\ref{section:conclusion} is reserved for concluding remarks.

\begin{figure*}
	\vspace{-10pt}
	\begin{center}
		\includegraphics[width=.7\textwidth]{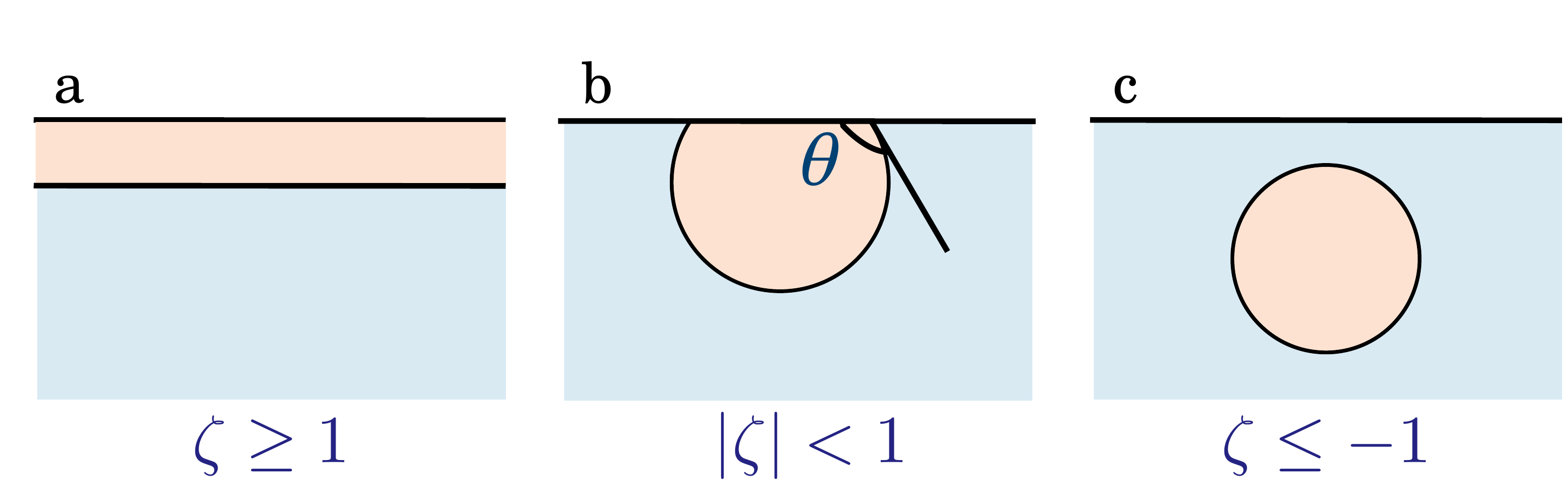}
		\caption{\label{fig:wetting}Different scenarios for surface freezing depending on the wetting parameter, $\zeta$: (a) For $\zeta\ge1$, the solid fully wets the free interface, and the freezing starts with the formation of a continuous solid layer at the surface. (b) For $|\zeta|<1$, the solid partially wets the free interface, and freezing starts with the formation of a spherical solid cap at the interface. (c) For $\zeta\le-1$, the solid does not wet the vapor-liquid interface, and freezing occurs exclusively in the bulk. }
	\end{center}
	\vspace{-10pt}
\end{figure*}

\section{\label{section:thermo} Thermodynamics of Surface Freezing}

The majority of thermodynamic models~\cite{TkachenkoPRL1996, TabazadehPNAS2002, ReissJPhysChemA2002, DjikaevJPhysChemA2008, QiuJACS2017, QiuCrystals2017} of surface freezing are  based on macroscopic arguments, which, as we will see later, can limit their applicability and predictive power. Such models are, however, very useful in providing a conceptual framework for understanding surface freezing.  Here, we present a simplified version of the model proposed in Refs.~\cite{TabazadehPNAS2002, ReissJPhysChemA2002, DjikaevJPhysChemA2008}, which is mathematically identical to the classical nucleation theory for heterogeneous nucleation~\cite{TurnballJCP1950} and is based on the following assumptions: (i) The crystal forming at the surface is structurally identical to the bulk crystal, (ii) All thermodynamic quantities, including  chemical potentials and surface tensions, are independent of the distance from or the curvature of the interface, (iii) Solid-fluid surface tensions are identical for different crystallographic planes of the solid, (iv) Enthalpic ($p\Delta V$) and line tension contributions to the free energy are negligible. With these assumptions, the free energy of formation for a crystalline body of volume $V_s$ at the free surface can be expressed as: 
\begin{eqnarray}
\Delta{G}_s &=& -\rho_s V_s\Delta\mu + A_{sl}\sigma_{sl}+ A_{sv}(\sigma_{sv}-\sigma_{lv})\label{eq:free_energy_general}
\end{eqnarray}
Here $\rho_s$ is the solid number density, $\Delta\mu=\mu_l-\mu_s$, is the thermodynamic driving force for freezing, $A_{sl}$ and $A_{sv}$ are the the solid-liquid and solid-vapor surface areas, and $\sigma_{sl}$, $\sigma_{sv}$ and $\sigma_{lv}$ are the solid-liquid, solid-vapor and liquid-vapor surface tensions, respectively. A material will undergo surface freezing if  $\Delta{G}_s$ is smaller than $\Delta{G}_b$, the free energy of formation of a crystalline nucleus of the same volume in the bulk for all $V_s>0$. In the context of this simple thermodynamic model, the occurrence of surface freezing, as well as the shape of the crystalline body will depend on  a dimensionless wettability parameter, $\zeta$, defined as:
\begin{eqnarray}
\zeta&:=&\frac{\sigma_{lv}-\sigma_{sv}}{\sigma_{sl}}
\end{eqnarray}
The most favorable condition for surface freezing is when $\zeta\ge1$, i.e.,~when the crystal \emph{fully} wets the liquid-vapor interface. In this case, a three-phase contact line becomes mechanically unstable, and  freezing always culminates in the formation of a spread frozen layer at the free interface (Fig.~\ref{fig:wetting}a). Note that in the fully wetting regime, the surface contribution to $\Delta{G}_s$ is always negative, and can thus compensate the unfavorable $\rho_s{V}_s\Delta\mu$ term at temperatures above $T_m^{\text{bulk}}$, resulting in the formation of a finite-thickness crystalline film at the free interface. This is the type of macroscopic surface freezing observed for $n$-alkanes, $n$-alcohols and some metallic alloys. 

The other extreme, $\zeta\le-1$, is when the crystal does \emph{not} wet the liquid-vapor interface. In this case, freezing can only start in the bulk (Fig.~\ref{fig:wetting}c) as any crystalline nucleus with a vapor-solid facet will be mechanically unstable, and will be pushed away from the interface. This wetting regime can also lead to a phenomenon known as \emph{surface pre-melting}~\cite{BienfaitEurophysLett1987, vanderGonSurfSci1990} in which a finite-thickness molten liquid layer will wet the solid-vapor interface at temperatures below $T_m^{\text{bulk}}$.

The third scenario is when $|\zeta|<1$, i.e.,~when the crystal \emph{partially} wets the vapor-liquid interface with a contact angle  $\cos\theta_c=\zeta$ (Fig.~\ref{fig:wetting}b). As outlined above,  the theoretical framework used for characterizing surface freezing in this regime is conceptually identical to the classical nucleation theory for heterogeneous nucleation~\cite{TurnballJCP1950}, in which the free energy of formation for a spherical crystalline cap of radius $r$ is given by:
\begin{eqnarray}
\Delta{G}_s(r) &=& f_c(\theta_c)\Delta{G}_b(r)\label{eq:free_energy_cap}
\end{eqnarray}
with $\Delta{G}_b(r)=\frac43\pi{r}^2(3\sigma_{sl}-r\rho_s\Delta\mu)$, the free energy of formation for a spherical nucleus of radius $r$ in the bulk, and $f_c(\theta_c)=\frac14(1-\cos\theta_c)^2(2+\cos\theta_c)$, the potency factor. Note that it is always more favorable to form a spherical cap of radius $r$ at the interface in comparison to a spherical nucleus of equal volume in the bulk, with the relative gain in free energy given by:
\begin{eqnarray}
\Delta{G}_s(r)-\Delta{G}_b(\bar{r}) &=& 4\pi{r}^2\sigma_{sl}\left[f_c(\theta_c)-f_c^{2/3}(\theta_c)\right]\notag\\
&\le&0
\end{eqnarray}
as $0\le f_c(\theta_c)\le 1$. However, $\Delta{G}_s(r)$ is a strictly increasing function of $r$ for $r\le r_c=2\sigma_{sl}/\rho_s\Delta\mu$, which means that a nucleation barrier of $\Delta{G}_s^* = 16\pi\sigma_{sl}^3f_c(\theta)/3\rho_s^2\Delta\mu^2$ needs to be crossed for freezing to proceed. Note that the free interface only decreases this barrier by a factor of $f_c(\theta_c)$ and does not eliminate it completely. It is worthwhile to mention that partial wettability can also lead to partial surface premelting, in which a partially wetting quasi-liquid layer emerges at the surface of ice at temperatures below $T_m^{\text{bulk}}$~\cite{ElbaumJCrystGrowth1993, MurataPNAS2016}.

This thermodynamic model provides a qualitative picture of the underlying thermodynamic forces that culminate in surface freezing. \textcolor{black}{A similar approach has been used to assess surface-induced heterogeneous nucleation propensity at flexible fluid-fluid and fluid-solid interfaces in terms of crystal-interface binding energies~\cite{QiuJACS2017, QiuCrystals2017}.}  There are, however, major issues that limit its applicability and predictive power. First of all, the assumptions outlined above are not usually satisfied in real experimental systems, and as will be discussed later, the predictions of this simple model have been shown to be inaccurate in computational studies of model systems~\cite{HajiAkbariFilmMolinero2014}. Secondly, it is extremely difficult to measure wettability accurately, especially in the deeply supercooled regime~\cite{JamesJCP2002}. This is particularly problematic in the partial wettability regime that is thought to lead to the interfacial enhancement of nucleation~\cite{KayAtmosChemPhys2003}.

\section{\label{section:experimental}Experimental Characterization of Surface Freezing}

\begin{figure}
\vspace{-10pt}
\begin{center}
	\includegraphics[width=.28\textwidth]{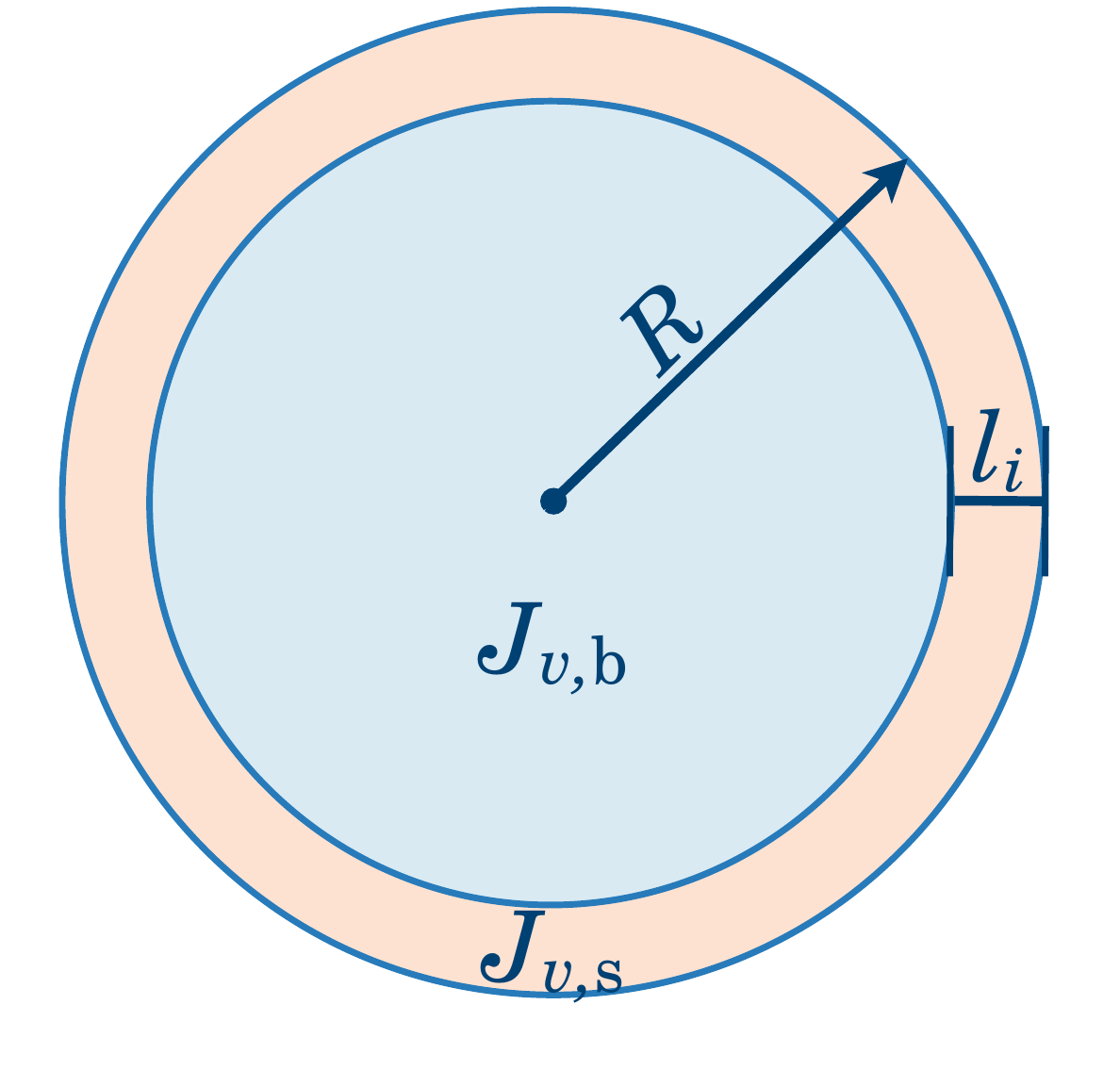}
	\caption{\label{fig:droplet}A droplet with radius $R$ and an $l_i$-thick interfacial region. The effective volumetric nucleation rates within the bulk and the interfacial region are $J_{v,\text{b}}$ and $J_{v,\text{s}}$, respectively.}
\end{center}
\vspace{-10pt}
\end{figure}

It is, in principle, straightforward to characterize the type of surface freezing that occurs at $T>T_m^{\text{bulk}}$ as it only requires establishing the existence of crystalline (or liquid-crystalline) order at the vapor-liquid interface. In other words, the thermodynamics of the confined system deviates so drastically from the bulk that it allows for a complete decoupling of structure and dynamics. There is a wide variety of experimental techniques that can be used for detecting liquid-to-solid structural transformation at the surface, and approaches such as surface tensiometry~\cite{WuScience1993, WuPRL1995, PfohlChemPhysLett1996, OckoPRE1997, GangPhysRevE1998}, X-ray scattering~\cite{WuPRL1995, OckoPRE1997, GangPhysRevE1998, ShpyrkoScience2006, ShpyrkoPhysRevB2007}, light scattering~\cite{FreylandJPhysCondMat2002}, polarized video microscopy~\cite{SwansonPRL1989},  elipsometry~\cite{PfohlChemPhysLett1996}, surface vibrational spectroscopy~\cite{SeflerChemPhysLett1995} and second harmonic and plasma generation spectroscopy~\cite{TurchaninPCCP2002} have been used to detect surface freezing in $n$-alkanes, $n$-alcohols, and metallic alloys.

The type of surface freezing that only involves an enhancement of nucleation kinetics close to a vapor-liquid interface is, however, far more difficult to detect and characterize experimentally, as its only distinction with respect to bulk nucleation is in the spatiotemporal distribution of nucleation events in the confined system, and not in the final outcome of the nucleation process. Considering the fact that the nucleation process involves the formation of a short-lived nanoscopic critical nucleus, any direct confirmation of surface-enhanced nucleation requires augmenting the structural assay distinguishing the liquid and the crystal with sufficient spatiotemporal sensitivity to measure the frequency and spatial distribution of isolated nucleation events. Unfortunately, this is not feasible with the existing traditional structural characterization techniques. There are only a limited number of sophisticated ultrafast scattering~\cite{GaffneyScience2007} and electron microscopy~\cite{PlemmonsChemMater2015, AdhikariACSApplMaterInter2017} techniques that can potentially achieve this goal in the future, but at present, they either cannot be  universally applied to a wide range of materials, or their sensitivity is yet to be improved to a point where phenomena such as surface-enhanced nucleation can be probed. 

\begin{figure*}
\vspace{-10pt}
\begin{center}
	\includegraphics[width=.9\textwidth]{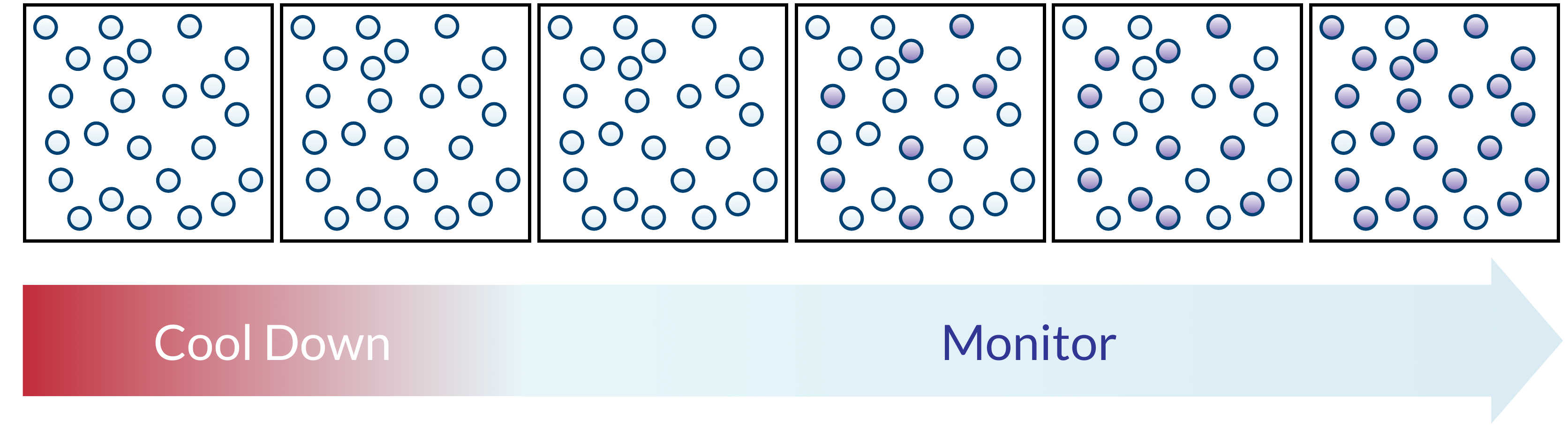}
	\caption{\label{fig:rate-measurement}A schematic representation of the approach used to measure homogeneous nucleation rates in experiments, which involves generating an ensemble of liquid droplets of the target material, cooling them down to the target temperature, and monitoring the fraction of the droplets that freezes by time $t$ using a structural or calorimetric assay. In this schematic, liquid and frozen droplets are depicted in light blue and light purple, respectively.}
\end{center}
\vspace{-10pt}
\end{figure*}

Considering these technical difficulties, surface-enhanced crystal nucleation has, by and large, been studied indirectly, by observing how the apparent volumetric nucleation rate, which is the average number of nucleation events per unit time per unit volume, scales with $l_c$ and temperature in confined geometries such as films and droplets. Consider, for instance, a droplet of radius $R\gg{l}_i$ (Fig.~\ref{fig:droplet}), with two distinct bulk-like (light blue) and interfacial (light orange) regions, and suppose that the volumetric nucleation rate within the bulk and interfacial regions are given by $J_{v,\text{b}}(T)$ and $J_{v,\text{s}}(T)$, respectively. The effective volumetric nucleation rate for the entire droplet can then be expressed as:
\begin{eqnarray}
\overline{J}_v&:=& J_{v,\text{b}}\left(1-\frac{3l_i}R\right) + J_{v,\text{s}}\frac{3l_i}{R}\approx J_{v,\text{b}}+J_{v,\text{s}}\frac{3l_i}{R}
\end{eqnarray}
If $J_{v,\text{b}}\overset{\gg}{\sim}{J}_{v,\text{s}}$, $\overline{J}_v$ will be independent of $R$, and so will be its scaling with temperature.   However, if $J_{v,\text{b}}\ll{J}_{v,\text{s}}$, $\overline{J}_v$ will depend on $R$, and there will a critical radius, $R_c=3l_iJ_{v,\text{s}}/J_{v,\text{b}}$ below which nucleation will be surface-dominated. Due to distinct local environments and nucleation barriers in the bulk and the interfacial region, it is reasonable to expect $J_{v,\text{b}}$ and $J_{s,\text{b}}$ to scale differently with temperature, in which case the scaling of  $\overline{J}_v$ with temperature will also be different for different $R$'s.

The dependence of $\overline{J}_v$ and its temperature scaling on $R$ can, in principle be used as a basis for determining whether a particular material would undergo surface freezing. In order to understand how to do that, we first need to discuss the experimental procedure for measuring homogeneous nucleation rates (Fig.~\ref{fig:rate-measurement}), which involves generating a large number of liquid droplets of the corresponding material, either in a vapor chamber, or as an emulsion within another liquid phase. Those droplets are then cooled down to the target temperature, and $F(t)$, the fraction of droplets frozen at time $t$, is estimated using an assay such as microscopy, scattering, spectroscopy or calorimetry. Assuming that the droplet size distribution is narrow, $F(t)$ is expected to follow the exponential distribution:
\begin{eqnarray}
F(t) & = & 1 - \exp[-\overline{J}_v\overline{V}t] \label{eq:Ft_mono}
\end{eqnarray}
with $\overline{V}$ the average volume of a droplet. As outlined above, surface-facilitated nucleation leads to a size-dependent rate and a size-dependent scaling of rate with temperature, and can therefore be probed by conducting rate measurements for droplets of different sizes, and observing their scaling with size and temperature. In the case of scaling with size, rates measured at a given temperature are expected to follow $\overline{J}_v(R)=A+B/R$ with $B>0$ implying surface-enhanced nucleation. The effect of temperature is, however, more complicated, since even if we assume the validity of classical nucleation theory, it is not at all clear how relevant quantities such as $\Delta\mu$ and $\sigma_{sl}$ change with temperature. Considering the narrow ranges of temperatures over which rate measurements can be conducted, it is usually reasonable to assume that the nucleation barrier is a linear function of temperature. In the case of surface-enhanced nucleation, it can be expected that the temperature dependence of the nucleation barrier, as inferred from $\overline{J}_v(T)$ measurements, will be vastly different for droplets of different sizes.

Despite having a sound theoretical basis, this approach is practically challenging, mostly due to the difficulty of generating microdroplets with narrow size distributions. In other words, the uncertainty in droplet size distributions might lead to large uncertainties in $B$, which could  make determining the occurrence of surface freezing practically impossible. This is particularly problematic if $R_c$ is towards the lower end of droplet sizes that can be generated experimentally for a particular material. Under such circumstances, the $R$ dependence of $\overline{J}_v$ (and its scaling with temperature) could be too weak to be accurately identified form the existing experimental data. Analyzing the temperature dependence of nucleation data has its own challenges, as it is not easy to control temperature in nucleation experiments, and even modest uncertainties in temperature can propagate to considerable errors in nucleation rates \textcolor{black}{due to their strong sensitivity to temperature~\cite{KoopZPC2004}}. \textcolor{black}{Finally, it is usually extremely challenging to conduct nucleation experiments under pristine conditions, i.e.,~in the absence of contaminants the presence of which can lead to unwanted heterogeneous nucleation~\cite{MurrayChemSocRev2012}.}

Considering these technical difficulties, it is generally difficult to prove or rule out surface-enhanced crystallization in materials, and as a result, most accounts of kinetic surface freezing are controversial. We will discuss these challenges and intricacies in the next section through our discussion of surface-enhanced freezing in liquid water.

\begin{figure*}
	\begin{center}
		\includegraphics[width=.9\textwidth]{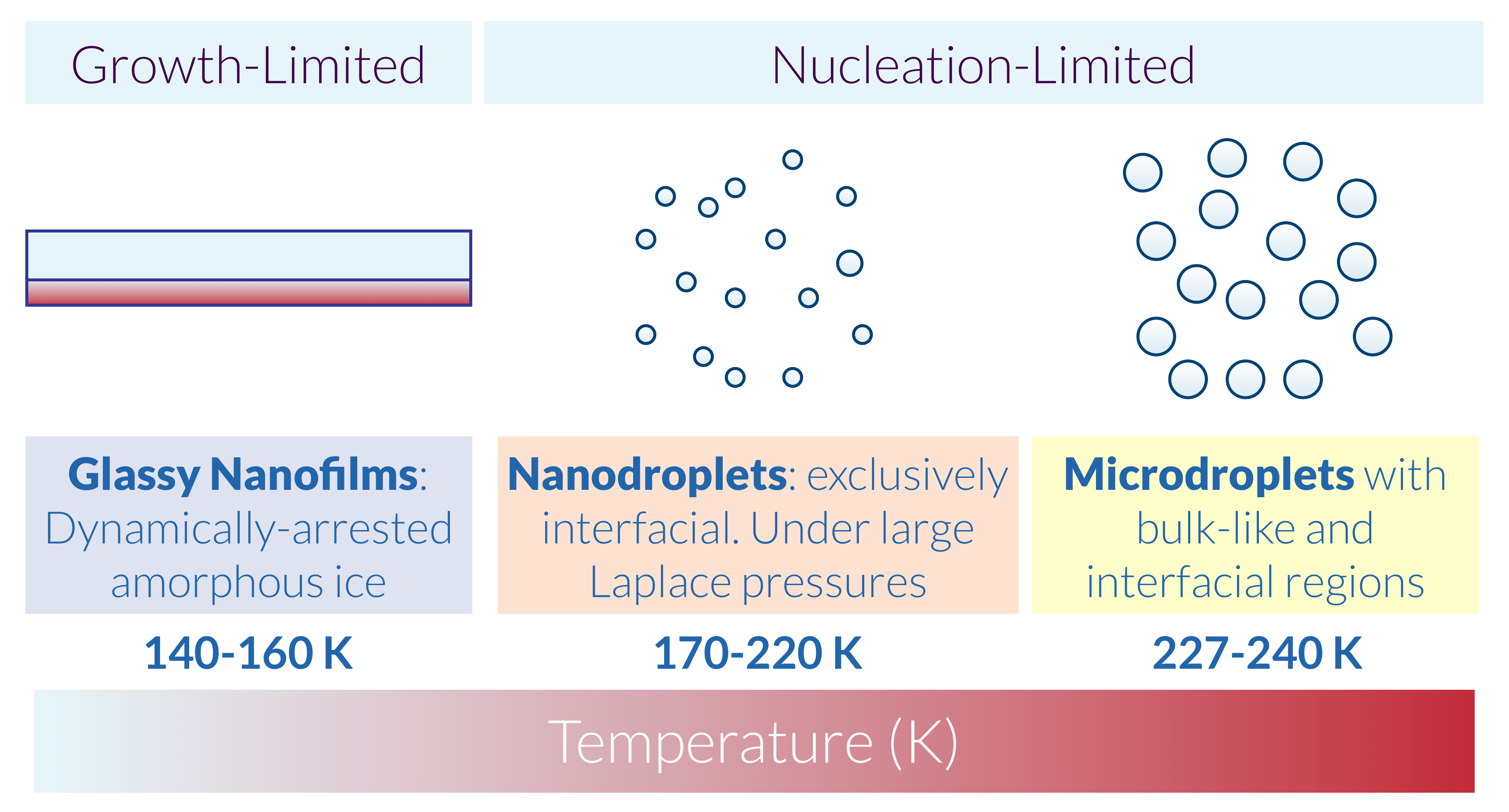}
		\caption{\label{fig:exploreSurfFreeze}Different approaches of probing surface-facilitated nucleation in water. Surface freezing can be either studied in the growth-limited regime (i.e.,~in nanofilms of amorphous ice), or in the nucleation-limited regime. The latter is done by both investigating the size and temperature scaling of homogeneous nucleation rates in microdroplets, or by inspecting the kinetics of nucleation in ultraconfined geometries such as nanodroplets. }
	\end{center}
\end{figure*}

\section{\label{section:water}Surface Freezing in Water}

\subsection{Experimental Work\label{section:water:experimental}}

At atmospheric pressure, liquid water becomes thermodynamically metastable with respect to ice at temperatures below $T_m^{\text{bulk}}=273$~K, but  is notoriously difficult to freeze in the absence of external insoluble entities. Indeed, ice formation can be avoided for  temperatures as low as 227~K~\cite{NilssonNature2014} and it has been recently estimated using molecular simulations and classical nucleation theory that homogeneous nucleation of ice is practically impossible at temperatures above 253~K~\cite{SanzJACS2013}, leaving heterogeneous nucleation as the only pathway for freezing at temperatures close to $T_m^{\text{bulk}}$. Homogeneous  nucleation rate measurements can thus only be conducted at deep supercoolings~\cite{KoopZPC2004}. A wide range of scattering~\cite{VonnegutJColloidSci1948, DuftACPD2004, StockelJPhysChemA2005, NilssonNature2014, NillsonJPCL2015}, microscopy~\cite{MurrayPCCP2010}, dilatometry~\cite{VonnegutJColloidSci1948, FoxNature1959} and calorimetry~\cite{TaborekPRB1985, CharoenreinThermochimActa1989, RiechersPCCP2013} techniques are utilized for detecting freezing in microdroplets placed in a variety of environments such as vapor \textcolor{black}{and expansion} chambers~\cite{DuftACPD2004, StockelJPhysChemA2005, BenzJPhotochemPhotobiolAChem2005, NilssonNature2014, NillsonJPCL2015}, reverse-phase oil emulsions~\cite{VonnegutJColloidSci1948, FoxNature1959, TaborekPRB1985, CharoenreinThermochimActa1989, RiechersPCCP2013} or on hydrophobic surfaces~\cite{MurrayPCCP2010}. As outlined in Section~\ref{section:experimental}, the most common way of proving surface-facilitated nucleation is to conduct rate measurements for droplets of different sizes and observe the scaling of rates with temperature. In addition to this conventional approach, the question of surface freezing in water can also be addressed using two other classes of experimental techniques (Fig.~\ref{fig:exploreSurfFreeze}). One possibility is to measure nucleation rates in ultra-confined geometries, such as nano-droplets, and assess the importance of surface freezing by extrapolating the measured rates to the corresponding bulk conditions. The second possibility is to measure freezing rates in the growth-limited regime, i.e.,~in nanofilms of amorphous ice, a glassy state of water obtained by rapidly quenching it to temperatures below 140~K~\cite{DebenedettiJPhysConsdensMatter2003}. Finally, more indirect anecdotal evidence for (or against) surface freezing  can be obtained from heterogeneous nucleation experiments as well.

\subsubsection{The original Report of Surface Freezing and Conventional Rate Measurements} 

The idea of surface-induced ice nucleation was first proposed in a series of papers by Tabazadeh~\emph{et al}~\cite{TabazadehPNAS2002, ReissJPhysChemA2002}. Their work was motivated by apparent discrepancies between the absolute values and the temperature scaling of earlier homogeneous nucleation rate measurements conducted for droplets of different sizes. By using a theoretical argument similar to the one discussed in Section~\ref{section:thermo}, they argued that any materials with $|\zeta|<1$ should undergo surface-enhanced nucleation. They then invoked the theoretical work of Cahn~\cite{CahnJCP1977}, who   predicted a wetting transition at temperatures sufficiently lower than the critical temperature, $T_c$, in single-component systems, and the experimental work of Elbaum~\emph{et al}~\cite{{ElbaumJCrystGrowth1993}} who observed that the quasi-liquid layer partially wets ice at coexistence, to conclude that water satisfies the partial wettability condition, and should therefore undergo surface-enhanced freezing.  They tested their hypothesis by re-analyzing earlier rate measurements by assuming that homogeneous ice nucleation in microdroplets is surface-dominated and therefore the nucleation time should scale with the surface and not the volume of the droplets. For measurements in which droplets were in contact with the vapor phase~\cite{DeMottJAtmosSci1990, KramerJCP1999, StockelJMolLiq2002}, the discrepancies became much smaller upon this re-analysis, which they interpreted to be a strong evidence in support of their surface freezing hypothesis. They even suggested that surface-enhanced nucleation is not limited to vapor-liquid interfaces, and can  occur at other water-fluid interfaces, such as some oil-water interfaces present in oil emulsion nucleation experiments. 

This work was initially met with some skepticism. First of all, Tabazadeh~\emph{et al} did not confirm the validity of the partial wettability condition for deeply supercooled water, and the only experimental work~\cite{ElbaumJCrystGrowth1993} that they cited as evidence was conduced close to the triple point, reporting a $\sigma_{sl}+\sigma_{lv}-\sigma_{sv}$ value three orders of magnitude smaller than the individual surface tensions. Some authors therefore posited the possibility that  $\sigma_{sl}+\sigma_{lv}-\sigma_{sv}$ could easily change sign at lower temperatures that are of relevance to homogeneous nucleation rate experiments~\cite{KayAtmosChemPhys2003}. Furthermore, the conceptual framework outlined in Section~\ref{section:experimental} is only exact when the constituent droplets are monodisperse. In reality, however, water droplets are always polydisperse, and this makes Eq.~(\ref{eq:Ft_mono}) inaccurate. Instead $F(t)$ will be given by: 
\begin{eqnarray}
F(t) &=& 1-\frac{\int f(R,t)V(R,t)e^{\left[-J_{v,\text{b}}V(R,t)-J_{v,\text{s}}l_iA(R,t)\right]t}dR}{\int f(R,t)V(R,t)dR}\notag\\&&\label{eq:Ft_poly}
\end{eqnarray} 
Here $f(R,t)$ is the droplet size distribution at time $t$ with $\int f(R,t)dR=1$. In order to distinguish bulk- vs.~surface-dominated nucleation, one needs to fit the existing $F(t)$ data to Eq.~(\ref{eq:Ft_poly}). As demonstrated in Ref.~\cite{SignorellPhysRevE2008}, however, the uncertainties in $F(t)$ are usually so large that a typical set of experimental data can be simultaneously described with the bulk- and the surface-dominated freezing scenarios.

Tabazadeh~\emph{et al}'s work was followed by a flurry of experimental activity with the aim of addressing some, if not all, of these technical difficulties~\cite{DuftACPD2004, LuApplPhysLett2005, EarleAtmosChemPhys2010, KuhnAtmosChemPhys2011, RzesankePCCP2012, RiechersPCCP2013}. One of the most important studies to follow was due to Earle~\emph{et al}~\cite{EarleAtmosChemPhys2010}, who developed a detailed microphysical model that accounted for heat and mass transfer effects, as well as droplet polydispersity, and used it to compute nucleation rates from freezing data obtained in vapor chambers. Also, more authors conducted rate measurements in vapor chambers, in order to obtain reliable nucleation data for droplets exposed to ambient air, as Tabazadeh~\emph{et al}~\cite{TabazadehPNAS2002, ReissJPhysChemA2002} only considered a limited number of such measurements in their analysis. Those later measurements revealed that surface-mediated nucleation can only become dominant for droplets smaller than a few micrometers in radius~\cite{EarleAtmosChemPhys2010, KuhnAtmosChemPhys2011}, and for larger droplets bulk nucleation is dominant~\cite{DuftACPD2004, RzesankePCCP2012, RiechersPCCP2013}. However, these results are still far from conclusive considering some of the problems outlined above. For instance, the work of Kuhn~\emph{et al}~\cite{KuhnAtmosChemPhys2011}, which so far offers the strongest evidence for surface-dominated ice nucleation in droplets smaller than 5~$\mu$m in radius, is based on a sophisticated microphysical model, and its conclusions can be fairly sensitive to the number of assumptions made in formulating that model. 

\subsubsection{Ultra-confined Geometries}

The importance of surface-dominated ice nucleation can also be potentially inferred from nucleation rate measurements in ultraconfined geometries, such as nanodroplets~\cite{HuangJPhysChem1995, MankaPCCP2012, BhabheJPhysChemA2013}. The main advantage of such studies is that nanodroplets lack a well-developed bulk region, and therefore any nucleation will be strongly impacted by the interface. Analyzing the rates obtained from such experiments, however, is not straightforward, as such droplets are under large Laplace pressures. (The Laplace pressure within a droplet of radius $r$ is given by $p_l=p_0+2\sigma_{lv}/r$, with $p_0$ the ambient pressure. A water droplet with a diameter of 10~nm is, for instance, under a Laplace pressure of $\sim$300~bar at 273~K.)  In principle, one can use classical nucleation theory to extrapolate such rates to pressures and temperatures that are relevant to conventional nucleation experiments. This, however, requires predicting how different thermodynamic and transport properties of supercooled water change within an experimentally inaccessible region of the metastable liquid phase diagram. Furthermore, pressure is known to change the structure of supercooled water~\cite{PooleNature1992, PalmerNature2014}, and it is not  at all clear whether ice nucleation at such high pressures will follow a mechanism commensurate with a one-step nucleation process predicted in the CNT formalism. Due to the fact that the volume of a nanodroplet is several orders of magnitude smaller than that of a microdroplet, freezing in nanodroplets occurs at lower temperatures, and at higher volumetric nucleation rates. Furthermore, the nucleation process culminates in the formation of stacking disordered ice \textcolor{black}{that is significantly more cubic~\cite{HuangJPhysChem1995, MankaPCCP2012, BhabheJPhysChemA2013, AmayaJPhysChemLett2017} than the stacking disordered ice formed in microdroplet experiments~\cite{MalkinPCCP2015, KoopJChemPhys2016}.}  The nucleation rates in nanodroplets are between 7-10 orders of magnitude higher than what would be obtained by extrapolating rate measurements in microdroplets at higher temperatures~\cite{NillsonJPCL2015}. However, this does not necessarily imply a contribution from surface freezing, with alternative scenarios having been proposed, such as a strong-to-fragile transition in supercooled water~\cite{NillsonJPCL2015}.

\subsubsection{Freezing in Amorphous Ice Nanofilms}

There have been numerous studies probing the freezing kinetics of amorphous ice nanofilms. Amorphous ice-- also known as amorphous solid water (ASW)-- is a glassy form of liquid water that can be obtained via a variety of pathways, including rapidly quenching the liquid to temperatures as low as 136~K, physically depositing water vapor onto a cold substrate~\cite{TammannZPhysChem1913}, or pressure-melting crystalline ice at low temperatures~\cite{MishimaNature1984}. Amorphous ice is the predominant form of ice in the interstellar space where temperatures are too low for water to crystallize~\cite{MitchellIcarus2017}. Understanding the kinetics and mechanism of amorphous ice crystallization is therefore crucial in mapping out astrophysical processes in the outer solar system. Consequently, the role of vapor-liquid interfaces in amorphous ice crystallization has been extensively studied~\cite{BackusPRL2004, BackusJCP2004, KondoJCP2007, YuanSurfSci2016, MitchellIcarus2017, YuanJCP2017}. The first major study was conducted by Backus~\emph{et al}~\cite{BackusPRL2004, BackusJCP2004}, who utilized reflection absorption infrared (RAIR) spectroscopy and temperature-programmed desorption (TPD) spectroscopy to distinguish bulk and surface crystallization, respectively, and concluded that ASW nanofilms freeze via a 'top-down` mechanism, in which freezing starts close to the vapor-glass interface.   Their final conclusion was, however, based on a detailed nucleation-and-growth model, and was not unequivocal. As a result, their findings were questioned in later publications~\cite{KondoJCP2007}. The most unequivocal evidence for surface freezing in ASW films was provided by Yuan~\emph{et al}~\cite{YuanSurfSci2016}, who preferentially placed isotopic layers of 5\% D$_2$O/95\%H$_2$O at different locations across a 1000-layer H$_2$O ASW nanofilm, and used RAIR spectroscopy to probe its crystallization. They observed that the isotopic layer crystallized faster when it was located closer to the vapor-ASW interface.  They later demonstrated that ASW films capped with decane freeze eight times more slowly than the films exposed to vapor~\cite{YuanJCP2017}. In another study, Mitchel~\emph{et al} demonstrated that high-porosity ASW films tend to freeze faster than their low-porosity counterparts, due to the presence of internal vapor-liquid interfaces within the porous material~\cite{MitchellIcarus2017}.

Among the three classes of approaches outlined above, amorphous ice freezing experiments provide more direct evidence for freezing at the surface. However, those findings must be treated with extreme caution. First of all, freezing of amorphous ice is a growth-limited process, and one can never rule out the possibility that faster freezing at the surface is merely due to faster dynamics at the surface. Note that faster diffusive dynamics does not equate to faster nucleation, especially when nucleation barriers are large, so the freezing kinetics data obtained in the growth-limited regime are not necessarily indicative of what would happen in the nucleation-limited regime. Secondly, due to the out-of-equilibrium nature of amorphous ice, its properties, including its freezing kinetics, can heavily depend on its processing history. This might explain part of the existing disagreement in the literature with regard to the role of a free interface. Finally, computer simulations of silicon have revealed that the ability of a vapor-liquid interface to enhance nucleation in its vicinity can depend on temperature and can completely disappear at temperatures closer to $T_m^{\text{bulk}}$~\cite{LiNatMater2009, LiJChemPhys2009}. Therefore, the enhancement of freezing at the vapor-glass interface of amorphous ice ($\sim$140-160 K), even if it affects the nucleation part, does not necessarily imply that the same  will be observed at higher and atmospherically relevant temperatures.

\subsubsection{Anecdotal Evidence for Surface Freezing}

\begin{figure}
	\begin{center}
		\includegraphics[width=.48\textwidth]{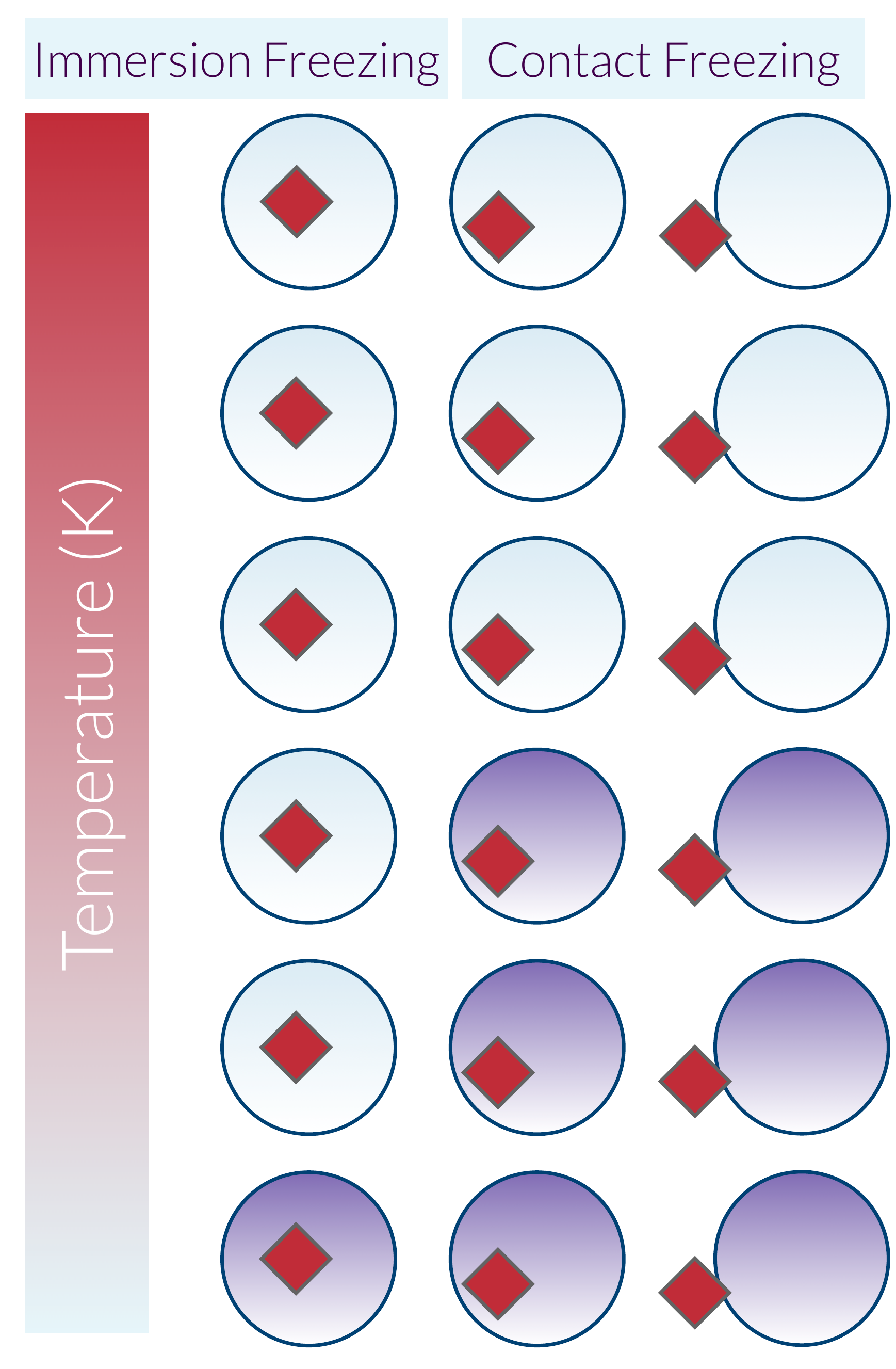}
		\caption{\label{fig:contact:freezing}Schematic description of immersion and contact freezing, with an external ice nucleating object depicted in dark red. Liquid and frozen droplets are shown in light blue and light purple, respectively. The occurrence of contact freezing is independent of whether the ice nucleating object approaches the interface from within, or collides with it from outside.}
	\end{center}
\end{figure}

In closing this section, it is worth mentioning a few experimental studies of heterogeneous nucleation that are relevant to the question of ice nucleation at vapor-liquid interfaces. Such studies are aimed at understanding a phenomenon known as \emph{contact freezing} (Fig.~\ref{fig:contact:freezing}), which involves heterogeneous ice nucleation in water droplets colliding with a dry ice nucleating agent (INA)~\cite{LadinoMorenoAtmosChemPhys2013}. Contact freezing has been shown to occur at rates considerably higher than \emph{immersion freezing} in which the INA is fully immersed within the droplet~\cite{ShawJPCB2005, DurantGeophysResLett2005}. This enhancement is observed even when the exogenous INA does not collide with the droplet from outside, but is instead approaching the free interface from within the droplet (inside-out contact freezing)~\cite{DurantGeophysResLett2005}. The observed enhancement in heterogeneous nucleation kinetics has therefore been attributed to the presence of a vapor-liquid interface, and not the transient effects arising from actual collisions. It has indeed been demonstrated that there is no preference for contact freezing to initiate at the actual three-phase contact line~\cite{GurganusJPhysChemLett2011} unless the INA surface has nanoscale texture~\cite{GurganusPRL2014}. Contact freezing is therefore generally regarded as anecdotal evidence for surface-induced homogeneous ice nucleation. However, there is no direct evidence for the assertion that surface-induced heterogenous nucleation is a sufficient condition for surface-induced homogeneous nucleation \textcolor{black}{(i.e.,~surface freezing)}.

\subsection{Computational Work\label{section:water:computational}}

As outlined in Section~\ref{section:water:experimental}, the existing experimental techniques lack the necessary spatiotemporal resolution to conclusively address the question of  surface freezing  in water. This has led to an increased interest in molecular simulations, which, by construction, can provide direct evidence for or against surface freezing. Conducting molecular simulations of ice nucleation has, however, its own challenges. Like any other molecule, simulating water requires identifying an empirical mathematic function known as a \emph{force field} or a \emph{model}, which describes the potential energy of the system as a function of the positions of the individual atoms. The multidimensional potential energy surface (PES) defined by the force-field guides the temporal evolution of the system, which, in the case of molecular dynamics (MD), is deterministic and involves integrating Newton's equations of motion. Utilizing a force-field is a convenient substitute to performing computationally expensive first principle calculations that can also be used for computing the PES of any given configuration. Over the years, a wide range of water force-fields with different levels of accuracy have been developed~\cite{JorgensenJChemPhys1983, GuillotJMolLiq2002, VegaPCCP2011}, including neural network-based force fields~\cite{HandleyJChemTheoryComput2009, MorawietzJPhysChemA2013}, classical polarizable~\cite{StillingerJChemPhys1978, BernardoJChemPhys1994, RickJChemPhys1994, ChenJPhysChemB2000, SternJChemPhys2001} and non-polarizable~\cite{StillingerJChemPhys1974, BerendsenPullman1981, JorgensenJChemPhys1983, BerendsenJPhysChem1987, NadaJCP2003, HornJChemPhys2004, VegaTIP4PiceJCP2005, VegaTIP4P2005, VegaPCCP2011} molecular force fields, and coarse-grained force-fields~\cite{MolineroJPCB2009, HadleyaMolSimulat2012, LobanovaMolPhys2014}. 
\begin{figure}
	\begin{center}
		\includegraphics[width=.46\textwidth]{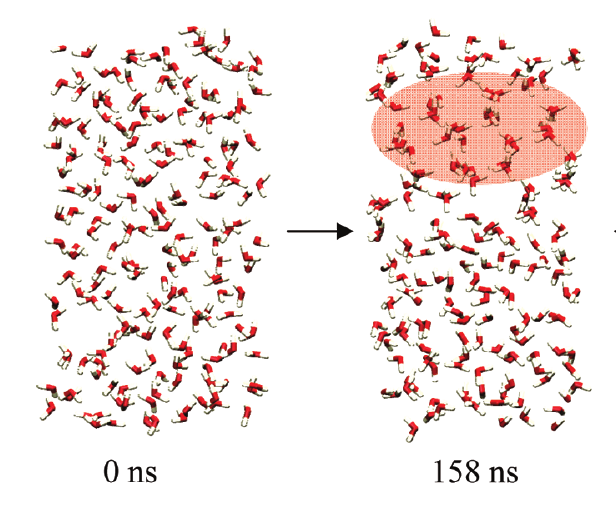}
		\caption{\label{fig:NE6} \textcolor{black}{(Reproduced with permission from \href{http://dx.doi.org/10.1021/jp9090238}{Pluha\'{r}ov\'{a},~et. al. J. Phys. Chem. C, 114, 7831 (2010)}, Copyright 2010, American Chemical Society)} Subsurface Freezing in MD simulations of freestanding thin films of the NE6 system at 250~K.}
	\end{center}
\end{figure}

\begin{figure}
	\begin{center}
		\includegraphics[width=.5\textwidth]{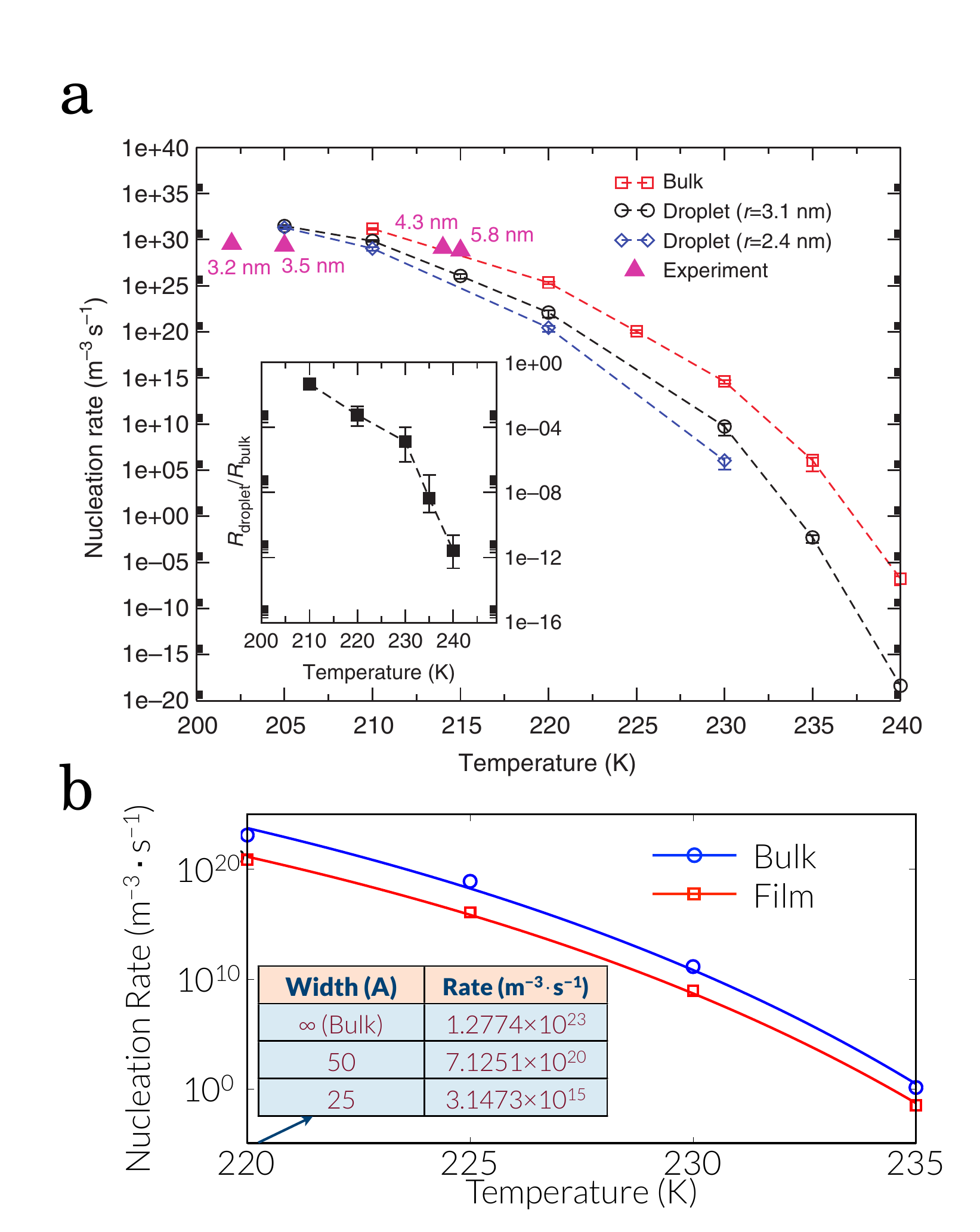}
		\caption{\label{fig:mW}Suppression of surface freezing in the mW system. (a) \textcolor{black}{(Reprinted by permission from Macmillan Publishers Ltd.: Nature Communications, (\href{http://dx.doi.org/10.1038/ncomms2918}{Li,~et al., Nat. Comm.~4, 1887 (2013)}), Copyright (2013))} Nucleation is always faster in the bulk than in nanodroplets. Experimental data are from Manka~\emph{et al}.~\cite{MankaPCCP2012}. The inset shows the ratio of the rate of homogeneous nucleation in 3.1-nm droplets over the nucleation rate in the bulk. (b) \textcolor{black}{(Reproduced from Ref.~\cite{HajiAkbariFilmMolinero2014} with permission from the PCCP Owner Societies)} Nucleation rates are always smaller in 5-nm-thick freestanding films of mW water than the bulk. The table corresponds to rate calculations  in films of different thicknesses at $T=220$~K.}
	\end{center}
\end{figure}

The predictive ability of a molecular simulation depends heavily on the accuracy of the utilized force-field. As expected, however, there is a direct relationship between the ability of a force-field to faithfully reproduce experimental properties of water, and the computational cost of using it. For the more accurate water models, such as polarizable models, even the simple task of structurally relaxing supercooled water can be prohibitively costly. For instance, the timescales accessible to state-of-the-art \emph{ab initio} MD simulations do not typically exceed 100~ps~\cite{NagataJChemPhys2016}, which is considerably shorter than the characteristic structural relaxation time of supercooled water computed from typical classical non-polarizable force-fields~\cite{HajiAkbariPNAS2015}. This is in addition to the activated nature of ice nucleation, which usually involves crossing large nucleation barriers that can, sometimes, be only overcome by employing advanced sampling techniques. Among the different classes of force-fields outlined above, ice nucleation has been successfully studied for coarse-grained and non-polarizable molecular models of water only,  and the more accurate polarizable and \emph{ab initio}-based models have, by and large, been off limits \textcolor{black}{due to prohibitively large computational costs of utilizing them in computational studies of nucleation}~\cite{MichaelidesChemRev2016}. This, in principle, can negatively impact the predictive ability of molecular simulations of surface freezing in water, considering the importance of polarizability in interfacial phenomena. Despite these limitations, molecular simulations can still be very valuable tools in uncovering the underlying physics of surface freezing, and how it relates to different thermodynamic, structural and dynamical features of an otherwise imperfect water model.

Computational studies of surface freezing date back to 2006 when \textcolor{black}{Vrbka}~\emph{et al}~\cite{JungwithJPCB2006, JungwirthJMolLiq2007} and Pluha\'{r}ov\'{a}~\emph{et al}.~\cite{ JungwirthJPhysChemC2010} conducted conventional MD simulations of ice nucleation in freestanding thin films of supercooled water using the NE6 model~\cite{NadaJCP2003}, a six-site non-polarizable molecular force-field specifically parameterized to reproduce the experimental thermodynamic properties of liquid water and ice around the melting temperature. These authors observed that most nucleation events started at the subsurface region, i.e.,~in the immediate vicinity of the vapor-liquid interface (Fig.~\ref{fig:NE6}). They attributed this behavior to interfacial ordering of water molecules, which led to the emergence of an electric field at the surface. Electric fields are known to induce homogeneous nucleation~\cite{PruppacherPureApplGeophys1973, KusalikPRL1994, KusalikJACS1996, BorzsakPhysRevE1997}. In other words, their argument for the facilitation of freezing at the interface was that it is a specific example of the already known phenomenon of electrofreezing. Orientational ordering of water molecules at a vapor-liquid interface has been observed for other molecular models~\cite{KathmannJACS2008, HajiAkbariPNAS2017}, and yet, none is known to undergo field-induced subsurface freezing at a free interface. This is even true for the models that spontaneously crystallize when a net electric field is applied to the entire system. \textcolor{black}{Furthermore, this idea of field-induced nucleation at the surface is not borne out by experimental evidence demonstrating that  negatively- and positively-charged droplets nucleate at the same rate as charge-neutral droplets~\cite{KramerJCP1999, RzesankePCCP2012}.} These findings might therefore be affected by strong finite size effects, as the simulation boxes used in Refs.~\cite{JungwithJPCB2006, JungwirthJMolLiq2007, JungwirthJPhysChemC2010} were too small in the directions parallel to the vapor-liquid interface ($L_x=1.35$~nm, $L_y=1.55$~nm, only 4-5 times the molecular diameter of $\sim$0.3~nm). In general, it is difficult to homogeneously nucleate ice in regular MD simulations of molecular models, and the handful of works reporting spontaneous ice nucleation in the absence of any external field or biasing potential, including the pioneering work of Matsumoto~\emph{et al}~\cite{Matsumoto2002} have never been reproduced in larger systems, and are  believed to reflect  finite size effects~\cite{SanzJACS2013, VegaJCP2014}.

Considering these difficulties, the next wave of computational activity did not arrive until after the introduction of the computationally efficient coarse-grained monoatomic water (mW) model~\cite{MolineroJPCB2009}, which is a re-parameterization of the widely known Stillinger-Weber potential for Group IV elements~\cite{StillingerPRB1985}. The mW potential has been very successful in reproducing the thermodynamic and structural properties of bulk water, and as a result, has gained considerable popularity in recent years. The first study of surface freezing using mW was due to Li~\emph{et al}, who used a path sampling technique known as forward flux sampling (FFS)~\cite{AllenFrenkel2006} to compute homogeneous ice nucleation rates in the bulk~\cite{GalliPCCP2011}, as well as nanodroplets~\cite{GalliNatComm2013} of mW water. They observed that nucleation in nanodroplets starts preferentially at the center of the droplets, and occurs at rates considerably lower than in the bulk (Fig.~\ref{fig:mW}a). This was qualitatively at odds with the reported behavior of the NE6 system, and highlights the challenges of using molecular simulations to address the surface freezing problem, as the final conclusion tends to depend on the utilized force-field. Another peculiar feature of Li~\emph{et al}.'s observation is its apparent inconsistency with their own earlier work on surface crystallization in silicon, another tetrahedral liquid, in which they concluded that surface-enhanced nucleation should occur for any material with a negatively-sloped solid-liquid coexistence line, i.e.,~with a liquid denser than the crystal~\cite{LiNatMater2009, LiJChemPhys2009}, and mW, despite satisfying this criterion, showed an apparent tendency to undergo bulk freezing. They explained this discrepancy by noting that water nanodroplets are under large Laplace pressures, and the thermodynamic driving force for crystallization decreases upon increasing pressure, which then leads to lower rates, and possibly alters the nucleation mechanism. However, Haji-Akbari~\emph{et al}~\cite{HajiAkbariFilmMolinero2014} and Gianetti~\emph{et al}~\cite{GianettiPCCP2016} used FFS to compute nucleation rates in freestanding thin films of mW and a few other mW-based tetrahedral liquids, and still observed lower nucleation rates in the film geometry for mW and another tetrahedral liquid satisfying the negatively-sloped coexistence line criterion (Fig.~\ref{fig:mW}b). A similar conclusion was reached in a work by L\"{u}~\emph{et al}, who used a mean-first passage time (MFPT) method~\cite{WedekindJChemPhys2007} to compute nucleation rates in films of different thicknesses~\cite{LuJPhysChemB2013}. Freestanding nanofilms have net zero curvature and are therefore not under Laplace pressure. As a result, the slower computed rates in film geometries cannot be explained by invoking higher pressures in the confined geometry. Furthermore, Haji-Akbari~\emph{et al}~\cite{HajiAkbariFilmMolinero2014} noted that the mW system \textcolor{black}{most likely} satisfies the partial wettability criterion~\cite{MolineroJACS2014}, and yet does not undergo surface freezing. This clearly demonstrates the inability of macroscopic arguments, including the model outlined in Section~\ref{section:thermo}, to predict phenomena as complex as surface freezing. 

\begin{figure}
	\begin{center}
		\includegraphics[width=.4\textwidth]{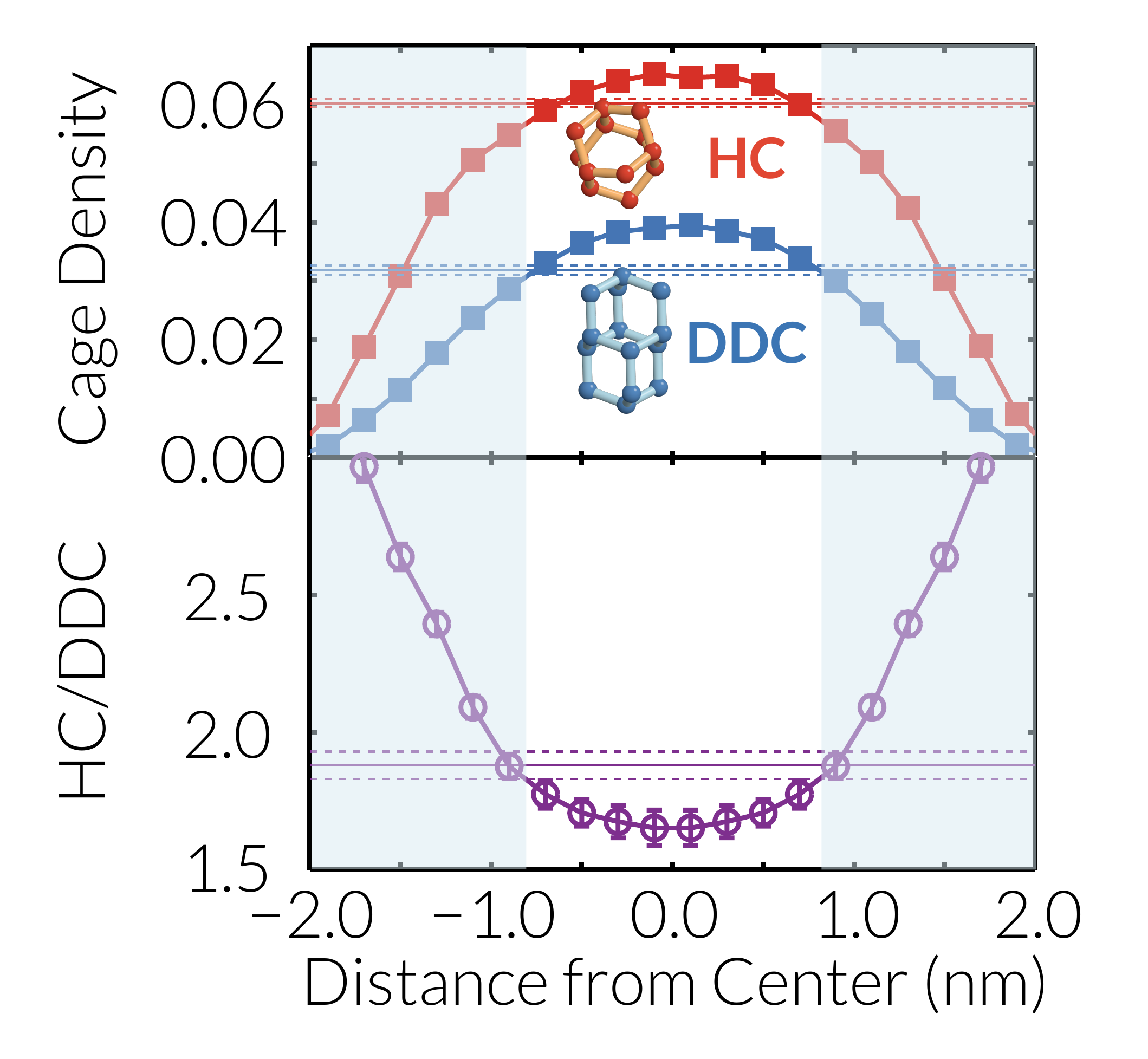}
		\caption{\label{fig:TIP4PIce} \textcolor{black}{(Reproduced from \href{http://dx.doi.org/10.1073/pnas.1620999114}{Haji-Akbari, Debenedetti,~Proc. Natl. Acad. Sci. USA~114, 3316 (2017)})}
		 The role of structural preference for cubic ice in the facilitation of ice nucleation in TIP4P/Ice nanofilms. HC and DDC number density (per nm$^3$) and HC/DDC ratio profiles are obtained from MD simulations of 4-nm-thick TIP4P/Ice films at 230~K. Horizontal lines correspond to the bulk cage densities and HC/DDC ratio at 230~K and 1~bar, with the dashed lines showing the error bars.  Nucleation starts at the center of the film that has lower-than-bulk HC/DDC ratios, and not in the shaded blue subsurface region. }
	\end{center}
\end{figure}

The qualitative difference between the NE6 and mW models created interest in accurately probing the surface freezing kinetics of more realistic molecular models, such as the TIP4P family. It is, however, almost impossible to homogeneously nucleate ice in MD simulations of  TIP4P-like systems, and even computing the rate and elucidating the mechanism using advanced path sampling techniques such as FFS was elusive for years. In 2015, Haji-Akbari~\emph{et al} developed and utilized~\cite{HajiAkbariPNAS2015} a coarse-grained variant of FFS to conduct the first direct calculation of the homogeneous ice nucleation rate for TIP4P/Ice~\cite{VegaTIP4PiceJCP2005}, one of the best existing non-polarizable classical molecular models of water. 
They then utilized the same method to compute the rate of homogeneous ice nucleation in a 4-nm-thick freestanding film under the same thermodynamic conditions~\cite{HajiAkbariPNAS2017}. Unlike the mW model, they observed an enhancement of nucleation in the film geometry. Interestingly, however, nucleation events started not at the immediate vicinity of the vapor-liquid interface (like in the NE6 system), but rather in a region of the film that  exhibited bulk-like behavior. Their detailed topological and structural analysis of the films, however, revealed  a preference for double-diamond cages (DDCs) over hexagonal cages (HCs) within the center of the film where freezing started (Fig.~\ref{fig:TIP4PIce}). DDCs and HCs are the topological building blocks of cubic and hexagonal ice, respectively. Cubic ice is an ice polymorph formed at deep supercoolings~\cite{MurrayNature2005}, and is a stacking variant of hexagonal ice, the thermodynamically stable form of ice at ambient pressures. Haji-Akbari~\emph{et al} had previously demonstrated that crystalline nuclei rich in DDCs grow more uniformly and are therefore more likely to contribute to the nucleation pathway~\cite{HajiAkbariPNAS2015}. They also conducted the same topological analysis for mW films, and observed no similar enhancement of cubicity. It therefore appears that this propensity for cubic ice formation, which is relevant to the microscopic mechanism of nucleation, also accurately predicts a water model's propensity to undergo surface freezing.  These findings are also interesting from a different perspective, as they demonstrate that certain structural features, such as cage number densities, decay to their bulk values over much larger length scales than what is usually considered a subsurface region. This is a key observation in understanding the nature of confinement and how it affects structural and dynamical properties of matter. The fact that non-decaying subtle structural features can impact the spatiotemporal distribution of nucleation events has also recently been observed in molecular simulations of silicon~\cite{LuJChemPhys2016}.

\section{Conclusions and Outlook\label{section:conclusion}}

It has been almost fifteen years since the idea of surface-enhanced ice nucleation was first proposed by Tabazadeh~\emph{et al}~\cite{TabazadehPNAS2002}. Since then, this idea has been thoroughly scrutinized through numerous experimental and computational investigations, which have resulted in more  indirect evidence in its support. Yet, we are still short of direct and unequivocal evidence for surface freezing, and more work is needed for a conclusive resolution of this conundrum. So far, the most direct experimental evidence for surface freezing has emerged from studies of crystallization in amorphous ice~\cite{BackusPRL2004, BackusJCP2004, YuanSurfSci2016, MitchellIcarus2017, YuanJCP2017}. However, considering the fact that freezing of glassy water is a growth-limited process, it is not clear whether those findings can conclusively imply a preference for nucleation at the interface. Rate measurements in small ($R<5~\mu$m) microdroplets tend to support the surface freezing hypothesis~\cite{EarleAtmosChemPhys2010, KuhnAtmosChemPhys2011}. Interpreting those results can, however, be non-trivial considering the sensitivity of  utilized microphysical models to the wide range of assumptions that have been made in their development. Studies of contact freezing (faster heterogeneous nucleation when an ice nucleating surface is close to the free interface) also provide some anecdotal evidence for the potential facilitating role of a free interface in nucleation~\cite{ShawJPCB2005, DurantGeophysResLett2005}. As sensible as it might seem, the assertion that faster heterogeneous nucleation close to a free interface would imply faster homogeneous nucleation as well is an unproven speculation. Computational studies of surface freezing are also far from conclusive and there is a divergence between atomistic models, which predict facilitation of freezing at a free interface~\cite{JungwithJPCB2006, JungwirthJMolLiq2007, JungwirthJPhysChemC2010, HajiAkbariPNAS2017}, and coarse-grained models, for which the contrary behavior is observed~\cite{GalliNatComm2013, HajiAkbariFilmMolinero2014, GianettiPCCP2016}.

Regarding the path forward, we can think of a few possibilities. On the experimental side, it will be worthwhile to develop better rate measurement techniques in order to better control temperature and droplet size distributions, and to measure ice nucleation rates in droplets smaller than a micrometer in diameter. Such experiments can provide more unequivocal evidence for surface freezing considering the fact that several studies discussed above suggest that surface freezing is likely to be dominant in submicron droplets. In addition, more effort should be invested into designing and optimizing better ultrafast scattering and electron microscopy techniques, which can provide more direct proof for surface-facilitated ice nucleation. This is particularly true about ultrafast electron microscopy techniques that are not currently usable for water. On the computational side, one possible area for exploration is the kinetics of ice nucleation in polarizable models, and the effect of polarizability in surface freezing. In addition, it is worthwhile to determine whether there are any thermodynamic or spectroscopic signatures that  correlate with surface-induced structural changes in  supercooled liquid nanofilms. Such signatures can potentially be used for probing the relevance of structural features observed in simulations in actual experimental systems.

\acknowledgements
P.G.D. gratefully acknowledges the support of the National Oceanic and Atmospheric Administration Cooperative Institute for Climate Science Award AWD 1004131; and of the Princeton Center for Complex Materials, a Materials Research Science and Engineering Center (MRSEC) supported by National Science Foundation (NSF) Grant DMR-1420541.


\bibliographystyle{apsrev}
\bibliography{References}

\end{document}